\documentstyle[twocolumn,aps,prl,epsf,psfig]{revtex}
\input{epsf}
\input epsf.sty
\begin{document} 

\title{$Ab$ $Initio$ Investigation of Collective Charge Excitations in MgB$_2$}
\author{Wei Ku$^1$, W.E.~Pickett$^1$, R.T.~Scalettar$^1$, and A.G.~Eguiluz$^2$}
\address{$^1$Physics Department, University of California, Davis, CA 95616}
\address{$^2$Department of Physics and Astronomy, The University of Tennessee,
Knoxville, TN 37996--1200, and}
\address{Solid State Division, Oak Ridge National Laboratory, Oak Ridge, TN
37831--6030}

\address{\mbox{ }} \address{\parbox{14cm}{\rm \mbox{ }\mbox{ } 
A sharp collective charge excitation is predicted in MgB$_2$ 
at $\sim2.5eV$ for $q$ perpendicular to the boron layers, based on 
an all--electron analysis of the dynamical density response 
within time--dependent density functional theory. This novel excitation, 
consisting of coherent charge fluctuation between Mg and B sheets, 
induces an abrupt plasma edge in the experimentally--observable reflectivity. 
The existence of this mode reflects the unique electronic structure of MgB$_2$, 
that is also responsible for strong electron--phonon coupling. 
By contrast, the acoustic plasmon, recently suggested to explain the high $T_c$, 
is not realized when realistic transition strengths are incorporated.
}}\address{\mbox{ }} \address{\mbox{ }}
\address{\parbox{14cm}{\rm \mbox{ }\mbox{ } PACS numbers: 74.70.Ad, 78.20.Bh, 78.20.-e
}} \maketitle

\narrowtext
The high critical temperature, $T_c\sim 40K$, 
of the newly discovered\cite{Nagamatsu}
conventional superconductor MgB$_2$ has attracted much 
attention\cite{An,Kortus,Kong,Voelker,Alexandrov,Yamaji,Bud,Martinho}.
A consistent picture within BCS 
theory\cite{Bardeen,Eliashberg,Scalapino} 
seems to be given by the qualitative analysis\cite{An,Kortus} 
based on the large hole density of states at the Fermi surface, 
provided by the boron $\sigma$--bands. 
On the other hand, other possible contributions to pairing\cite{Yamaji,Bud,Bickers} 
have not been totally ruled out, since a quantitative analysis of $T_c$, 
via realistic treatment of screened electron--electron and electron--phonon 
interactions\cite{Kong}, has yet to be performed. 
The insight into the behavior of the cuprate superconductors 
resulting from consideration of the spin susceptibility 
suggests that it might be especially interesting to examine 
the possibility of unusual collective charge excitations in MgB$_2$ with 
such realistic treatments. In particular, a calculation of the 
dynamical density response would give detailed information on electronic 
excitations, which control most of the macroscopic properties, including 
the dielectric function and thereby superconductivity\cite{Scalapino}.

Even within phonon-mediated theories, 
there currently exists large speculation of the value of the Coulomb pseudopotential 
$\mu^*$, from 0.02 to 1\cite{Voelker,Alexandrov}. 
An intriguing proposal\cite{Voelker} has recently suggested that 
an anomalously small $\mu^*$ may originate from the existence of an acoustic plasmon 
(AP)\cite{Pines,Frohlich},
which develops from two characteristic charge carriers in MgB$_2$: 
``heavy" holes of B $\sigma$--bands and ``light" electrons and holes of B $\pi$--bands. 
This is an important issue, as the existence of an AP can drastically alter 
the dynamical electronic screening and 
may even provide an additional pairing channel\cite{Frohlich}. 
Thus, it is vital to re-examine the existing 
tight--binding result via a more realistic investigation that 
takes effects of the crystal potential fully into account.

In this Letter we present results of an $ab$ $initio$, all--electron, calculation 
of the dynamical density-response function of MgB$_2$ within 
the framework of time-dependent density functional theory (TDDFT)\cite{Petersilka}. 
Our key result is the prediction of a sharp collective mode 
whose origin is the strong coherent charge fluctuation
between parallel sheets of B and Mg. This novel charge excitation, found 
to reside within the [0002] zone-boundary gap, embodies a remarkable signature 
of the electronic structure of MgB$_2$, and results in a dramatic 
change in the reflectivity near $2.5 eV$ for $q$ perpendicular to the boron layers. 
It thus explains naturally the strong increase in the Raman efficiency 
reported very recently\cite{Martinho}.
The impact of this mode on the Coulomb pseudopotential will be briefly addressed. 
In addition, with the matrix elements
that control the transition probability fully included, the recently 
proposed acoustic plasmon is not realized in our results, as a consequence of
weak intraband transition of the ``heavy carriers". 
Finally, comments on the proposed two-band-type superconducting instabilities 
will be given to further illustrate the importance of including 
realistic transition strengths.

Within TDDFT, the dynamical density response function 
$\chi(\vec x,t;\vec x \,',t')=\delta \rho(\vec x,t)/
\delta v_{ext}(\vec x \,',t ')$
with density $\rho$
and external potential $v_{ext}$, 
can be obtained through the following integral 
equation\cite{Petersilka}:
\begin{equation}
\chi=\chi^{KS}+\chi^{KS}(v+f_{xc})\chi,
\end{equation}
where $\chi^{KS}(\vec x,t;\vec x \,',t')$
is the response function for ``non--interacting" Kohn--Sham (KS) 
electrons, $v(\vec x - \vec x \,')$
is the Coulomb interaction, and 
$f_{xc}(\vec x,t;\vec x \,',t')=\delta v_{xc}(\vec x,t)/
\delta \rho(\vec x \,',t ')$
accounts for dynamical 
exchange--correlation effects (where $v_{xc}$ is the time--dependent 
exchange--correlation potential in TDDFT.) Working 
in Fourier space, $\chi^{KS}$
can be calculated with the KS eigenenergies, $\epsilon_{\vec k,n}$ 
and the corresponding eigenstates, $|\vec k,n\rangle$, 
of the ground state\cite{Blaha}:
\begin{eqnarray}
&&\chi_{\vec G,\vec G \,'}^{KS}(\vec q,\omega)=
{1 \over V} \sum_{\vec k}^{BZ} \sum_{n,n'}
{f_{\vec k,n}-f_{\vec k+ \vec q,n'} \over
\epsilon_{\vec k,n}-\epsilon_{\vec k+ \vec q,n'} + \hbar(\omega+i0^+) }
\nonumber \\
\times && \langle \vec k,n | e^{-i(\vec q+\vec G) \cdot \vec x}
| \vec k+ \vec q,n' \rangle
\langle \vec k+ \vec q,n' | e^{i(\vec q+\vec G \,') \cdot \vec x}
| \vec k,n \rangle,
\end{eqnarray}
where $\vec G$ is a vector of the reciprocal lattice, $n$ is a band index, 
the wave vectors $\vec k$ and $\vec q$ are in the first 
Brillouin zone, and $V$ is the 
normalization volume. Equation~(1) is then numerically solved as a matrix 
equation\cite{crystalfields}.

We stress that this formalism is rigorous. Even though a formal 
connection between the KS eigenenergies and the 
quasi--particle energies has not been established, the physical 
meaning of the KS band structure (including the empty states) is 
to be realized through its contribution to the density fluctuation in $\chi^{KS}$. 
The only physical approximation introduced in this work is the 
functional form of $v_{xc}(\vec x,t)$,
for which the adiabatic local density approximation 
(ALDA)\cite{Petersilka} is chosen. Since only the long wavelength limit is of 
interest in the present work, we further drop $f_{xc}$ from Eq.~(1), as its 
strength (within ALDA) cannot compete with the singular $v$ 
without sizable crystal local--field effects\cite{crystalfields}. 
In this limit, the dielectric function $\epsilon=(1+v\chi)^{-1}$ can be visualized
directly from the single--particle transitions of the 
KS particles, with the relation $\epsilon=1-v\chi^{KS}$.

Numerically, in order to resolve the features in the narrow energy range 
where the AP may occur, the broadening parameter, 
commonly employed\cite{Eguiluz} in place of $0^{+}$ in Eq.~(2), 
needs to be very small($\sim10meV$). This in turn requires
a dense $k$--mesh, to carefully sample the momentum--energy phase space near the
Fermi surface. The results 
presented in this work correspond to a uniform mesh of $20\times20\times60$. 
In addition, summing the narrow $\delta$--functions is avoided by 
first evaluating Eq.~(2) on the imaginary frequency axis, followed by 
analytic continuation via Pad\'{e} approximants\cite{Ku}.

\begin{figure}[hbt] 
\unitlength1cm \begin{picture}(8.0,4.8) \put(-0.3,-1.3)
{\psfig{figure=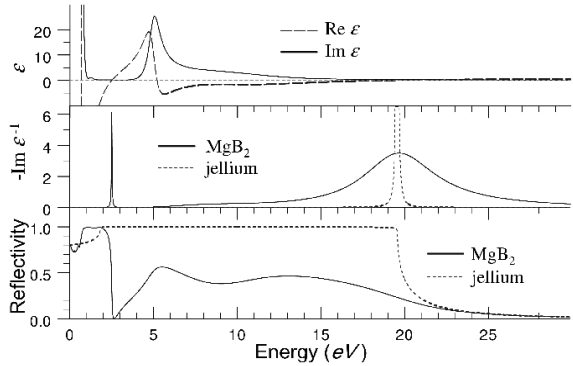,height=6.4cm,width=8.5cm,angle=-90}}
\end{picture} 
\vskip 0.2cm
\caption{
Calculated loss function (upper panel), dielectric function (middle panel), 
and reflectivity (lower panel), as well as the jellium 
counterparts (dashed lines) for $q = 0.12$\AA$^{-1}$, along the $c$--axis.
}
\end{figure}

The calculated dielectric function and corresponding 
loss function ($-$Im$\,\epsilon^{-1}=-v\,$Im$\chi$) along the $c$--axis, 
which should be directly probed by inelastic scattering experiments of 
electrons or x-rays\cite{Fink}, are shown over a broad energy range in Fig.~1. 
The wide structure\cite{finestructure} at $\omega \sim 20eV$ in the loss function 
is the conventional 
plasmon corresponding to a density of $\sim 8$ electrons per unit 
cell $(r_s \sim 1.8 a_0$). 
It suffices to note the correspondence between the near 
vanishing of the dielectric function and the peak line shape of the loss 
function, illustrating the collective nature of this over--damped mode.

The most striking feature of the loss function shown in Fig.~1 is 
the sharp peak that occurs at $\sim2.45eV$. 
The origin of this peak is evidently the prominent absorption at $5eV$ in Im$\,\epsilon$, 
whose physics will be addressed shortly. 
Indeed, the strength of this charge fluctuation (notice the vertical scale of $\epsilon$) 
is large enough to generate, through screening, an additional zero in Re$\,\epsilon$ 
thereby inducing a new collective mode. 
The extremely long lifetime ($\Delta \omega < 10meV$) of this mode 
is a consequence of both the strong dynamical screening 
in the vicinity of its energy, as reflected in the large slope 
of Re$\,\epsilon$\cite{Ku,Sturm1}, and the lack of particle--hole decay channels, 
courtesy of the large energy gap to be discussed below.

This additional collective excitation has a dramatic physical impact
on the reflectivity, as shown in the bottom panel of Fig.~1. 
When possessing enough energy to excite this mode, 
the incident light suddenly penetrates the surface of the 
material with no reflection, much more abruptly than the case of normal metal with a plasmon.

The reduction of reflectivity also provides a natural qulitative explanation 
for the large enhancement of the Raman effeciency\cite{Sarma} 
in a recent $2.41eV$ Raman scattering study\cite{Martinho} of the E$_{2g}$ phonon, 
as more photons are allowed to enter the surface to interact with the system.

\begin{figure}[hbt] 
\unitlength1cm \begin{picture}(8.0,5.35) \put(-0.5,-0.75)
{\psfig{figure=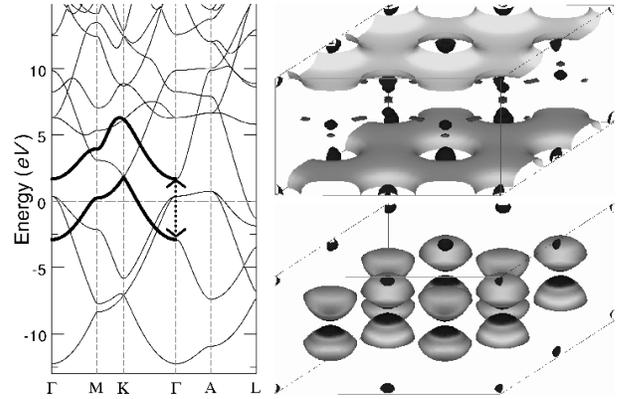,height=6.4cm,width=8.5cm,angle=-90}}
\end{picture} 
\vskip 0.2cm
\caption{
Calculated band structure and density. The arrows in the band 
structure (left panel) indicate the large $5eV$ gap between parallel 
bands. The isosurfaces of density for states marked by the upper/lower 
arrows are shown in the upper/lower right panel, which contains $2\times2\times1$
unit cells with Mg in the corners.
}
\end{figure}

Examination of the band structure, shown in the left panel of Fig.~2, 
reveals that the $5eV$ structure in Im$\,\epsilon$ 
originates mainly from the interband transitions between parallel bands. 
These transitions are facilitated by the strong [0002] crystal potential, 
which reflects the layered structure of this system and 
gives a large $5eV$ gap (indicated by the arrows) at the $\Gamma$ point. 
Intriguingly, as shown in the right panel of Fig.~2, 
where the real--space charge distribution of states marked by the down and up arrows 
is illustrated, the states involved in the transitions reside in different planes 
and correspond to a B $p_z$ $\pi$--state and 
a layered conducting state of Mg $s$--symmetry. 
Evidently, the above-mentioned additional mode consists of 
$coherent$ charge fluctuations between B and Mg sheets, 
under the strong dynamical screening of low-energy intraband transitions.\cite{Sturm2} 

Note that this mode is of very different nature from the $d$-plasmon observed in 
a few post-transition metals.  
The $d$-plasmon consists of $direction$-$insensitive$ collective $dipole$-like oscillation 
of the $localized$ electrons, 
originating from transitions of flat $d$--$bands$ to a $wide$ range of final states.  
Thus, the corresponding absorption in Im$\,\epsilon$ is usually 5-10 times smaller than the current case.

The wave vector dependence of the new collective mode, controlled by the competition 
between dynamical intraband screening and interband charge fluctuation, 
is shown in Fig.~3, in which the surface illustrates underlying 
single-particle transitions given by Im$\,\epsilon(q//[0001],\omega)$.
The first branch at $\omega\sim0.5eV$ of the low--energy ridge of the surface 
corresponds to the intraband charge fluctuation of the heavy carriers 
($p_{x,y}$ $\sigma$--bands) to be discussed below, and the second branch to the light 
carrier ($p_z$ $\pi$--bands), while the high-energy ridge demonstrates the $5eV$ 
interband transition. In the valley between these two main structures, where 
the phase space for single-particle excitation is closed due to the 
[0002] crystal potential, lies the weakly dispersed, induced collective 
mode (marked by dots), which eventually decays into electron--hole pairs 
of the light carriers at $q \sim 0.6$\AA$^{-1}$. 

\begin{figure}[hbt] 
\unitlength1cm \begin{picture}(8.0,5.15) \put(-0.5,-1.1)
{\psfig{figure=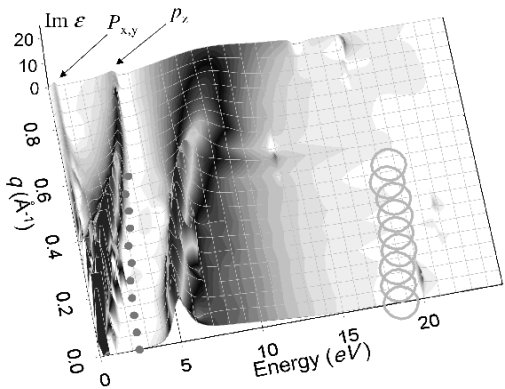,height=6.7cm,width=9.0cm,angle=-90}}
\end{picture} 
\vskip 0.2cm
\caption{ 
Dispersion of collective modes and the single--particle decay channel. 
The dispersion of the induced/nominal collective mode is denoted by 
dots/circles until it becomes difficult to identify the peak energy of the plasmon. 
The surface illustrates Im$\,\epsilon(q,\omega)$ for $\vec q//[0001]$.
}
\end{figure}

It is natural to wonder about the impact of this 
additional mode on the superconductivity, through reduction of the 
Coulomb pseudopotential, $\mu^*$.  Within the static approximation adopted by 
previous authors\cite{Eliashberg,Scalapino}, this new mode is found to have almost no 
effect on the static screened interaction, due to its very sharp line 
shape and small spectral weight compared to the conventional
plasmon. In fact, the static limit of our resulting screened interaction 
is almost identical to that obtained from the homogeneous electron gas, 
despite its highly anisotropic energy dependence.  
One thus would expect a $\mu^{*}$ of ``narmal" magnitude, 
if it is evaluated from the screened interaction\cite{Lee}.  
However, due to the strong $k$--dependence of the Fermi surfaces, 
even the very idea of treating the screening through a single Coulomb pseudopotential 
may not be adequate (c.f.~Eq.~(2.26) of Ref.~\cite{Scalapino}).

Now we address the proposed \cite{Voelker} AP with 
calculated dielectric function for small momentum/energy, shown in Fig.~4. 
The two main peak structures in Im$\,\epsilon$ are related to the unusual 
characteristic of the Fermi surface\cite{Kortus}. The peak of lower energy 
corresponds to the intraband charge fluctuation of the heavy carriers, 
while the main peak results from the light carriers. More 
specifically, for a small momentum transfer $q$ along the 
$c$--axis, the cylindrical $p_{x,y}$ $\sigma$--hole Fermi surfaces\cite{Kortus} 
contribute to smaller energy (heavier mass), compared to the three-dimensional
$p_z$ $\pi$--hole and $\pi$--electron Fermi surfaces. 
This double--peak structure is precisely what led to the suggestion of AP\cite{Voelker}, 
as the plasmon energy of the heavy carriers might be greatly screened 
by the light carriers and reduced to the upper edge of low-energy peak, 
whose energy increases linearly with small $q$. 

\begin{figure}[hbt] 
\unitlength1cm \begin{picture}(8.0,3.9) \put(-0.4,-2.15)
{\psfig{figure=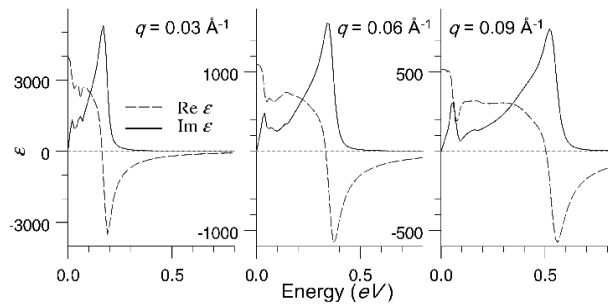,height=6.4cm,width=8.5cm,angle=-90}}
\end{picture} 
\vskip 0.2cm
\caption{
Calculated dielectric function of MgB$_2$, for small $q$'s along the $c$--axis. 
The low/high-energy peaks of 
Im$\,\epsilon$ correspond to charge 
fluctuation of the heavy/light carriers. Note the large difference in 
the spectral weight and that Re$\,\epsilon \gg 1$ 
between these two main structures.
}
\end{figure}

However, this simple picture neglects electron--hole decay channels and 
the relative strength (or spectral weight) of these two carriers. As 
shown in Figure 4, for small $q$ there is a large supply of charge fluctuations 
(Im$\,\epsilon \gg 1$) involving the light carriers at the upper edge of 
the low-energy peak. As a result, any collective mode of this 
energy would rapidly decay into electron--hole pairs (Landau damping) and 
have a negligible lifetime. In fact, our analysis suggests that a gap 
in the light--carrier bands at the Fermi energy is needed to allow an 
AP to exist. Furthermore, in this energy range, Re$\,\epsilon$ 
is considerably larger than unity, reflecting the fact that the spectral weight 
from the heavy carriers is insignificant compared to the strength of the 
screening from the light carriers. In short, the potential 
acoustic plasmon is totally screened out and should not contribute to $\mu^*$.
(The low energy intra-$\sigma$-transitions may still lead to non-adiabatic corrections 
to the high energy, small $q$ phonons, in particular the $E_{2g}$ modes, 
which are most important for superconductivity.)

This kind of interplay between these two structures in $\epsilon$ cannot be faithfully 
estimated by just counting the phase space, as performed in Ref.~\cite{Voelker}
with parameterized tight--binding bands. Instead, a careful evaluation 
of the corresponding transition probability, controlled by the matrix 
elements in Eq.~(2), is necessary. The commonly made approximation of 
unit probability risks 
promoting unphysical charge fluctuations in the calculation, especially in the 
case of multi--band models\cite{Adolph}.

For a similar reason, the recently proposed two-band-type superconducting instability\cite{Yamaji}
in MgB$_2$ is also not supported by our results. For $q \sim (0,0,\pi/c)$, 
where the nesting between B $\pi$--electron and 
$\pi$--hole Fermi surfaces becomes ``perfect", there is no 
apparent enhancement at zero energy in our dielectric functions 
($q \sim 0.9$\AA$^{-1}$ in Fig.~3), contrary to the large interband polarizability 
obtained in Ref.~\cite{Yamaji}. 
Even though the formalism adapted in this work is 
different from that in Ref.~\cite{Yamaji}, 
based on the experience gathered in the 
past years\cite{Eguiluz,Ku},
we are confident that our results should not suffer from qualitative error for a 
simple $sp$ system like MgB$_2$. Thus, the enhancement of interband pair 
scattering in Ref.~\cite{Yamaji} is an artifact of omitting the transition 
probability (c.f. Eq.~(9) of Ref.~\cite{Yamaji}) as discussed above.

In conclusion, a novel sharp collective mode, involving 
coherent charge fluctuation between B and Mg sheets, is predicted. This 
additional collective mode has a great impact on the optical properties 
of the material -- suddenly switching off the reflectivity in the corresponding energy. 
It thus naturally explains the strong increase in the reported Raman efficiency at $2.41eV$, 
by increasing chance of interaction. 
By contrast, the recently suggested AP is not supported by 
careful evaluation of the dynamical density response functions, 
with the full information of all--electron KS band structure. 
The weak charge fluctuation from heavy carriers is totally screened out by 
the light carriers. Thus, AP is effectively ruled out as a possible mechanism 
that brings an anomalously small Coulomb pseudopotential, $\mu^*$. 
Our results further demonstrate the importance of 
inclusion of realistic crystal potential effects, through the band 
dispersions (gaps) and matrix elements, in the calculation of charge fluctuation 
for systems with non-spherical Fermi surfaces.

This work was supported by DOE Grant DE--FG03--01ER45876 and an 
Accelerated Strategic Computing Initiative grant through Lawrence 
Livermore National Laboratory.  
A.G.E. and R.T.S. acknowledge support from NSF and NERSC.


\begin{references}
\bibitem{finestructure} 
Fine structures in the plasmon line shape (e.g.~a dip in the center) 
are suppressed for the ease of visualization.
\end{references}
\end{document}